\documentclass[11pt]{amsart}
\usepackage{mathrsfs}
\usepackage[english]{babel}
\usepackage{fontenc}
\usepackage{graphicx}

\usepackage{color}
\usepackage{amssymb}
\usepackage{amsmath}

\usepackage{graphicx}

\newtheorem{theorem}{Theorem}[section]

\newtheorem{prop}[theorem]{Proposition}
\newtheorem{lem}[theorem]{Lemma}

\newtheorem{defi}[theorem]{Definition}
\newtheorem{remark}[theorem]{Remark}

\newenvironment{rem}{\begin{remark} \rm}{\end{remark}}

\def\la{\lambda}

\newcommand{\CN}{\mathcal N}

\newcommand{\CQ}{{\cal Q}}

\newcommand{\tr}{\mathrm{tr}\,}

\def\dpt#1#2{{\frac{\partial #1}{\partial #2}}}
\def\dncoo{Darboux-\Nij\ coordinates}
\def\omnman{$\omega N$ manifold}

\def\Nij{Nijenhuis}

\def\pd#1{\frac{\partial}{\partial#1}}

\def\endpf{\begin{flushright}$\square$\end{flushright}}

\newcommand{\Gr}{{{\mathcal{G}(\la)}}}
\newcommand{\Ek}{{{\mathcal{E}(\la)}}}

\newcommand{\mb}[1]{{\mathbf{#1}}}
\newcommand{\bs}[1]{{\boldsymbol{#1}}}

\def\parpo#1#2{\{#1,#2\}}
\def\parp#1{{\parpo{\cdot}{\cdot}}_#1}
\def\ddd#1#2{\displaystyle{\frac{\partial #1}{\partial #2}}}

\def\la{\lambda}        
\newcommand{\dsl}[1]{{\displaystyle{#1}}}
\newcommand{\RR}{{{\mathbb{R}}}}

\begin{document}

\selectlanguage{english}

\title[Canonical spectral coordinates for the CM-system]{Bi-Hamiltonian geometry and canonical spectral 
coordinates for the rational  Calogero-Moser system}



\begin{abstract}
We reconsider the rational Calogero-Moser system from the point of view of bi-Hamiltonian geometry. 
By using geometrical tools of the latter, we explicitly construct set(s) of spectral canonical coordinates, 
that is, complete sets of Darboux coordinates defined by the eigenvalues and the eigenvectors of the Lax matrix.
\end{abstract}

\author{Gregorio Falqui and Igor Mencattini}

\date{\today}

\address{Gregorio Falqui, Dipartimento di Matematica e Applicazioni, Universit\`a di Milano Bicocca,
Via R.Cozzi, 55, I-20126 Milano, Italy}

\email{gregorio.falqui@unimib.it}

\address{Igor Mencattini, ICMC-USP Universidade de S\~ao Paulo, Avenida Trabalhador
Sao-carlense 400 Centro, CEP: 13566-590, S\~ao Carlos, SP, Brazil}

\email{igorre@icmc.usp.br}

\maketitle

\tableofcontents
\section{Introduction}\label{Intro}
The rational Calogero-Moser system 
(CM-system from now on) is a dynamical system consisting of $n$-particles constrained to move on a line and pairwise interacting under the influence of a potential proportional to the inverse squared distance of the particles.
Its Hamiltonian is
\begin{equation}
H(x,p)=\frac12 \sum_{j=1}^{n}{p_j}^2+g
\sum_{i<j}\frac1{(x_i-x_j)^2}.\label{eq:HamCM}
\end{equation}
The Hamiltonian dynamical system described by \eqref{eq:HamCM} has a long and rich history. In the case $n=3$ it already appeared in the work of Jacobi \cite{Ja1866}, while the general case started to be studied systematically only at the end of the sixties of the last century, within the broader scheme of the completely integrable classical and quantum systems. In \cite{Ca69}, Calogero solved the $n$-particle quantum analogue of this dynamical system and in \cite{Ca71}  he conjectured the complete integrability of the classical Hamiltonian~\eqref{eq:HamCM}. This conjecture was proved to be true by Moser in \cite{Mo75}, where, after rewriting the equations of motion defined by \eqref{eq:HamCM} in a Lax form, explicit solutions were also obtained. 

The so called {\it projection method} was introduced in \cite{OlshPer76}  to find an alternative way to integrate the CM-system. The relation of this method with the theory of the momentum map and with the Lax representation discovered by Moser was discussed by Kazhdan, Kostant and Sternberg, KKS from now on, in \cite{KKS78}, where the complete integrability of the Hamiltonian \eqref{eq:HamCM} was proved using a symplectic reduction technique. Finally, in \cite{Wo83} it was proved that this system is super-integrable. Nowadays the CM-system plays the role of a unifying concept which stands at the crossroad of many areas of mathematics, from representation theory, see for example \cite{Et06}, to non-commutative geometry, see for example \cite{Ginz01}, to mention just a few.

Along with the inverse scattering method, based on the existence of a Lax representation for the equations of motion, another important mathematical framework that was brought to the light in the modern theory of integrable systems was that of  bi-Hamiltonian geometry.  While for most integrable systems a sort of dictionary between the two settings was  established, the bi-Hamiltonian setting for CM-system escaped a similar systematization for quite a long time.  Indeed, the bi-Hamiltonian formulation of the CM-system was obtained  by Magri and Marsico in \cite{MM96}  in the framework of a detailed local study of a special class of bi-Hamiltonian manifolds,  that is, the class of symplectic manifolds endowed with a suitable $(1,1)$-tensor (see Section \ref{S2.1} for detailed definitions\footnote{It is still not known whether the trigonometric and the elliptic Calogero-Moser systems admit a bi-Hamiltonian description.}). 

To the bi-Hamiltonian representation  of the CM-system was given a broader meaning in \cite{BFMOP10}, where the Magri-Marsico brackets were derived by a double process of Hamiltonian reduction from a suitable bi-Hamiltonian structure on the cotangent bundle of the Lie algebra $\mathfrak{gl}(n)$, thus making contact with the classical KKS group-theoretical reduction procedure~\cite{KKS78}.

A drawback of this approach is that in the standard  canonical coordinates $(x_j,p_j)$ the second Poisson structure cannot be explicitly computed, since the transformation that gives  the Magri-Marsico coordinates in terms of the physical ones proves to be too difficult to be inverted. 
Indeed this was done only for the case of $n=3$, in \cite{AR12}.
This possibly parallels the fact that also the $R$-matrix representation of the  Poisson brackets for the CM-system is quite awkward, since it involves \cite{AT93} a {\em dynamical} $R$-matrix, that is, an $R$-matrix whose elements depend on the phase-space variables.

The goal of the present paper is the close up this circle of ideas. Using  the setting of \cite{BFMOP10}, we shall at first introduce a set of canonical coordinates whose ``first half" is composed by the eigenvalues of the Calogero-Moser Lax matrix. Then we shall discuss the meaning of these coordinates in the framework of the algebro-geometric  setting for integrable systems.\\

\emph{Outline of the paper:} in Section 2 and Section 3 we shall review those notions of
 bi-Hamiltonian geometry to be used in the core of the paper. More precisely, in Section 2  we shall recall the notion of Darboux-Nijenhuis coordinates, together with a generalization thereof to be termed {\em magnetic} Darboux-Nijenhuis coordinates. Then, we will expand on some techniques related to the notion of Darboux-Nijenhuis coordinates, already introduced in \cite{FP03}, that will be thoroughly used in the paper. In Section 3 we shall review, following   \cite{BFMOP10}, both the properties of a special bi-Hamiltonian structure on $T^*\mathfrak{gl}(n)$ and its reduction to the phase-space of the rational CM-system. Finally, in Section 4, we shall show how this setting  can be fruitfully applied to the CM-system to provide explicit formulas for spectral canonical coordinates (whose expression was suggested in a talk in 2009 at Cambridge by E.K. Sklyanin, but, to the best of our knowledge, has not appeared in the literature yet). We close the paper showing how some (known) features of the rational CM-system can  simply and directly be recovered in the formalism herewith presented.

\section{$\omega N$-manifolds and their local geometry}\label{S2.1}

The basic definitions of bi-Hamiltonian geometry needed in this paper stem from the notion of $\omega N$-manifold.  A $\omega N$-manifold is a triple $(M,\omega,N)$ where $(M,\omega)$ is a symplectic manifold (that is, $\omega$ is a closed non-degenerate $2$-form on $M$), and $N$ is a $(1,1)$- tensor,  i.e. an endomorphism of the tangent bundle to $M$, satisfying the following {\em compatibility} conditions:
\begin{description}
\item[i)] For all vector fields $X,Y$
\begin{equation}
\label{nn1}
\omega(NX,Y)+\omega(NY,X)=0.
\end{equation}
\item[ii)] The $2$-form $\omega^\prime$ implicitly defined in the preceding item as 
\begin{equation}
\label{nn2}
\omega^\prime(X,Y):=\omega(NX,Y)
\end{equation}
is closed.
\end{description}
It can be shown that the closure condition  $d \omega^\prime=0$ is equivalent to the vanishing of the Nijenhuis torsion $T_N$  of the $(1,1)$-tensor $N$
 \begin{equation}\label{nn3}
  T_N(X,Y):=[NX,NY]-N\big([NX,Y]+[X,NY]-N[X,Y]\big).
 \end{equation}
It is customary to call $(1,1)$-tensors satisfying (\ref{nn3}) {\em Nijenhuis} tensors. Note that every $\omega N$-manifold  is a \emph{bi-Hamiltonian} manifold. Indeed,  
recalling that in the symplectic picture the Hamiltonian vector field $X_f$ corresponding to $f\in C^{\infty}(M)$ can be defined via
\begin{equation}
\label{nn4}
\omega(X_f, Y)=-\langle df, Y\rangle\, ,
\end{equation}
one can see that, setting 
\begin{equation}\label{nn5}
{\parpo{f}{g}}_0:=\omega(X_f,X_g),\qquad {\parpo{f}{g}}_1:=\omega^\prime(X_f,X_g)\, ,
\end{equation}
the compatibility conditions (\ref{nn1},\ref{nn2}) above imply that $\parp{0}$ and $\parp{1}$ form a pair of compatible Poisson brackets, that is,  for all $\la\in\RR$, $\parp{0}+\la\parp{1}$ is a Poisson bracket. The pair so defined is called the pair of Poisson brackets associated with the $\omega N$-manifold $(M,\omega,N)$.

Conversely, if   $P_0$ and $P_1$ are two compatible Poisson tensors on $M$, with, say, $P_0$ invertible, then $(M, P_0^{-1}, P_1 P_0^{-1})$ is a $\omega N$-manifold. Here we consider Poisson tensors as maps from the cotangent bundle to the tangent bundle to $M$, defined via
\begin{equation}
\label{nn5-1}
{\parpo{f}{g}}_\ell=\langle df, P_\ell\, dg\rangle, \quad \ell=0,1.
\end{equation}
We remark for further use that in this case the adjoint of the recursion tensor $N$ (to be considered as an endomorphism of the cotangent bundle  $T^*M$) is simply given by $N^*= P_0^{-1}P_1$.

Let $(M,N,\omega)$ be a  $\omega N$-manifold of dimension $2n$. Then:

\begin{defi}
A set of local coordinates $(x_i,p_i)$ on $M$ is called a set of {\it Darboux-Nijenhuis (DN) coordinates}  for $(M,N,\omega)$ if:
\begin{enumerate}
\item They are canonical coordinates for the symplectic form $\omega$, i.e. $\omega=\sum_{i=1}^ndy_i\wedge dx^i$,
\item They reduce the Nijenhuis tensor $N$ to the diagonal form
\begin{equation}
 N = \sum_{i=1}^{n}\lambda_i \left(\pd{x_i} \otimes dx_i + 
\pd{p_i} \otimes dp_i\right).\label{eq:N}
\end{equation}
\end{enumerate}
\end{defi}
If we consider the pair of Poisson brackets defined on $(M,\omega, N)$  as in Eq. \eqref{nn5} then 
we get the following fundamental Poisson brackets
\[
\{x_i,p_j\}_0=\delta^i_{j}\,\,\text{and}\,\,\{x_i,x_j\}_0=0=\{p_i,p_j\}_0,\,\forall\, i,j=1,\dots,n,
\]
while
\[
\{x_i,p_j\}_1=\lambda_i\delta^i_{j}\,\,\text{and}\,\,\{x_i,x_j\}_1=0=\{p_i,p_j\}_1,\,\forall\, i,j=1,\dots,n.
\]

\begin{remark}
Formula \eqref{eq:N} implies that all the eigenvalues of $N$ have even multiplicity.
\end{remark}

A $2n$-dimensional $\omega N$-manifold $(M,N,\omega)$ is called {\em semi-simple} at $m\in M$ if there exists a neighborhood of $m$ where its Nijenhuis tensor admits $n$ distinct eigenvalues $\lambda_1,\dots,\lambda_n$.  It is called {\em regular} at $m$ if there exists a neighborhood of $m$ where the eigenvalues are functionally independent.

Let us define  $I_k=\frac{1}{2\, k}\operatorname{tr}(N^k)$. Using the vanishing of the Nijenhuis torsion of $N$, one can prove the so called {\em Lenard recursion relations}: 
\begin{equation}
 N^{\ast}dI_{i}=dI_{i+1},\,\forall i=1,\dots n.\label{eq:lenard}
\end{equation}
Furthermore, as a consequence of the compatibility between $\omega$ and $N$, it can be also proven that
\begin{equation}
 \{I_i,I_j\}_0=\{I_i, I_j\}_1= 0,\,\forall i,j=1,\dots,n.\label{eq:inv}
\end{equation}
Then differentiating both sides of the identities
\begin{equation}
kI_k=\lambda^k_1+\cdots +\lambda_n^k,\,\forall k=1,\dots,n\label{eq:symmpol}
\end{equation}
and using the Lenard relations, see  \eqref{eq:lenard}, one arrives at
\[
 N^{\ast}d\lambda_k=\lambda_kd\lambda_k,\,\forall k=1,\dots,n.
\]
Using the invertibility of \eqref{eq:symmpol} and the \eqref{eq:inv}, one gets that  the eigenvalues of $N$ are in involution with respect to both  brackets, that is, 
\[
 \{\lambda_i,\lambda_j\}_0= \{\lambda_i,\lambda_j\}_1=0,\,\forall i,j=1,\dots,n.
\]
In other words, in a neighborhood of a point $m\in M$ where the $\omega N$-manifold is regular and semi-simple, the eigenvalues of the Nijenhuis tensor provide {\em one-half} of a set of DN-coordinates. Furthermore, this set of coordinates can be always completed to a full set of DN-coordinates, as stated, more precisely, in the following proposition.

\begin{prop}\label{pro:DNcoor}
 In a neighborhood of a point where a $2n$-dimensional $\omega N$-manifold is regular and semi-simple, given $\lambda_1,\dots,\lambda_n$ the eigenvalues of $N$, it is possible to find by quadrature $n$-functions $\mu_1,\dots,\mu_n$ such that the $2n$-functions $(\lambda_1,\dots,\lambda_n,\mu_1,\dots,\mu_n)$ define a set of DN-coordinates.
\end{prop}

\begin{rem}\label{rem:imp}
Under the assumptions made in Proposition \ref{pro:DNcoor} on the $\omega N$-manifold, the $2n$-functions $(\lambda_1,\dots,\lambda_n,\mu_1,\dots,\mu_n)$ form a set of (local) DN-coordinates if and only if:
\begin{enumerate}
\item[(1)] $N^*d\mu_i=\la_id\mu_i,\,\,\forall i=1,\dots,n$, and
\item[(2)] $\{\mu_i,\mu_j\}_0=0,\quad \{\la_i,\la_j\}_0=0$ and $\{\la_i,\mu_j\}_0=\delta_{ij},\,\forall i,j=1,\dots,n$.
\end{enumerate}
One can prove that, if the conditions in $(1)$ are fulfilled, then $(2)$ can be replaced by the $n$ conditions:
\begin{equation}
\{\la_1+\cdots +\la_n,\mu_i\}=0,\,\,\forall i=1,\dots,n,\label{eq:hamvf}
\end{equation}
which do not require the explicit evaluation of the $\la_i$'s, but only of their sum, that is, one-half of the trace of $N$. Indeed,  the condition expressed in Formula \eqref{eq:hamvf} is equivalent to
\[
Y(\mu_i)=1,\,\forall i=1,\dots,n,
\]
where 
\begin{equation}
Y=-P_0dI_1=\sum_{j=1}^ n\frac{\partial}{\partial \mu_j}.\label{eq:vech}
\end{equation}
\end{rem}

\subsection{Magnetic DN-coordinates}\label{mdn}
Let $(M,N,\omega)$ be a $2n$-dimensional $\omega N$-manifold, and let  $P_0$ the Poisson tensor defined by $\omega$.
Along with the DN-coordinates above recalled, we will be interested in a further class of coordinates, 
to be called for short {\em Magnetic-Darboux-Nijenhuis (MDN-) coordinates} . 
They were already used in ~\cite{MM96} as an intermediate step towards the definition of \dncoo, but will acquire a more prominent role in our discussion of the CM-system.

Given a set of DN-coordinates on a $\omega N$-manifold $M$, a set of MDN-coordinates can be defined by 
a $\la$-{\it dependent shift} in the momenta $\mu_i$. More precisely, a set of MDN-coordinates $(\tilde\la_1,\dots,\tilde\la_n,\tilde\mu_1,\dots,\tilde\mu_n)$ can be  obtained from a set of DN-coordinates $(\la_1,\dots,\la_n,\mu_1,\dots,\mu_n)$ by the (local) transformation: 
\begin{equation}\label{MDN-coo}
\tilde{\la}_i=\la_i,\quad  \tilde{\mu}_i=\mu_i+f_i(\la_1,\cdots, \la_n),\,i=1,\dots,n,
\end{equation} 
where $f_1,\dots,f_n$ are smooth functions satisfying the conditions:
\begin{equation}
 \frac{\partial f_i}{\partial\lambda_j}=\frac{\partial f_j}{\partial\lambda_i},\,\forall i,j=1,\dots,n.\label{eq:intcond}
\end{equation}
As soon as each of the $f_i$'s is not a {\em separated} function of the corresponding $i$-th coordinate $\la_i$, it is immediate to see that the MDN-coordinates are not DN-coordinates. However, they have the following properties:
\begin{prop}\label{pro:magco}
Setting $P_1=NP_0$, and denoting, as above, with $\{\cdot,\cdot\}_1$ and $\{\cdot,\cdot\}_0$  the corresponding Poisson brackets, then: 
 \begin{enumerate}
  \item[1.] The coordinates  \eqref{MDN-coo} are canonical coordinates for $\{\cdot,\cdot\}_0$  and
  \item[2.] They reduce the Nijenhuis adjoint tensor in the (block-type) {\em Jordan} form:
  \begin{equation}
   \label{JN-form}
   N^*=\left(\begin{array}{cc} \Lambda& \beta\\
              0&\Lambda
             \end{array}\right)
  \end{equation} 
where $\Lambda$ is the diagonal matrix with  entries $(\la_1,\dots,\lambda_n)$, while $\beta$ is related with the transformation (\ref{MDN-coo}) by:
\begin{equation}
 \beta_{ji}=\lambda_j\frac{\partial f_i}{\partial\lambda_j},\,\forall i,j=1,\dots,n \label{eq:beta}.
\end{equation}
\item[3.] The $P_1$--fundamental Poisson brackets  are:
\begin{equation}\label{p1pb}
{\parpo{\tilde\la_i}{\tilde\la_j}}_1=0, \quad {\parpo{\tilde\la_i}{\tilde\mu_j}}_1=\delta_{ij}\tilde\la_j,\quad \quad {\parpo{\tilde\mu_i}{\tilde\mu_j}}_1=B_{ij}.
\end{equation}
where $B_{ij}=\beta_{ij}-\beta_{ji},\,\forall i,j=1,\dots,n.$
\end{enumerate}
\end{prop}
{\bf Proof.} Part $1.$ follows by computing the elementary Poisson brackets $\{\widetilde{\lambda_i},\widetilde{\lambda_j}\}$, $\{\widetilde{\lambda_i},\widetilde{\mu_j}\}$ and $\{\widetilde{\mu_i},\widetilde{\mu_j}\}$ and by noticing that these last ones are zero for all $i,j=1,\dots,n$ if and only if $f_1,\dots,f_n$ fulfill the conditions \eqref{eq:intcond}.
To prove part $2.$ it suffices to note that since $(\lambda_1,\dots,\lambda_n,\mu_1,\dots,\mu_n)$ is a set of DN-coordinates then it holds
\begin{equation}\label{eigenNstar}
 N^{\ast}d\lambda_i=\lambda_id\lambda_i,\,\quad
 N^{\ast}d\mu_i=\lambda_id\mu_i,\,\forall i=1,\dots,n\, .
\end{equation}
Then,  $N^{\ast}d\widetilde{\lambda_i}=\widetilde{\lambda_i}d\widetilde{\lambda_i},\,\forall i=1,\dots,n,$ still holds, while
\begin{equation}\begin{split}
 N^{\ast}d\widetilde{\mu_i}&=N^{\ast}d\mu_i+N^{\ast}df_i
 =\lambda_id\mu_i+\sum_{j=1}^n\frac{\partial f_i}{\partial\lambda_j}N^{\ast}d\lambda_j\\
 &=\lambda_id\mu_i+\sum_{j=1}^n\frac{\partial f_i}{\partial\lambda_j}\lambda_jd\lambda_j
 =\lambda_id\mu_i+\sum_{j=1}^n\beta_{ji}d\lambda_j.
\end{split}
\end{equation}
Finally, the proof of part $3.$ is obtained by a direct computation, recalling that $\{f,g\}_1=\omega(NX_f,X_g)$, for all $f,g\in C^{\infty}(M)$.
\endpf 

\begin{remark} 
{\rm The  geometrical interpretation of the coordinates defined in \eqref{MDN-coo} is obtained as follows.  Let $m\in M$ be a point where $(M,N,\omega)$ is regular and semi-simple and let  $(\lambda_1,\dots,\lambda_n,\mu_1,\dots,\mu_n)$ be a set of DN-coordinates defined on a suitable neighborhood $V$ of $m$. Let us identify $V\simeq\mathbb R^{2n}\simeq T^{\ast}\mathbb R^n$  endowed with its canonical symplectic structure, $\omega=\sum_{i=1}^nd\lambda_i\wedge d\mu_i$. Let $\pi:V\rightarrow\mathbb R^n$ be the canonical projection, defined, at the level of the DN-coordinates, by $\pi(\lambda_1,\dots,\lambda_n,\mu_1,\dots,\mu_n)=(\lambda_1,\dots,\lambda_n)$. 
Then \eqref{MDN-coo} represents a diffeomorphism of $T^{\ast}\mathbb R^n$, defined by the {\it translation along the fibers} of the canonical projection $\pi$. In this framework, one should think of the functions $(f_1,\dots,f_n)$ as the components of a $1$-form $\alpha$, i.e. $\alpha=\sum_{i=1}^nf_id\lambda_i$. In particular, \eqref{eq:intcond} is equivalent to the closure of $\alpha$ and  Point $1$ of Proposition \ref{pro:magco} is nothing more than the well known statement that given a manifold $Q$, the diffeomorphism $t_{\alpha}:T^{\ast}Q\rightarrow T^{\ast}Q$, defined by $t_{\alpha}(\xi)=\alpha+\xi$, where $\alpha\in\Omega^1(Q)$, is a symplectomorphism {\it if and only if} $d\alpha=0$. Finally, since $d\alpha=0$, there locally there exists $S\in C^{\infty}(\mathbb R^n)$ such that $f_i=\frac{\partial S}{\partial\lambda_i}$, for all $i=1,\dots,n$, so that one can write 
\[
B_{ij}=(\lambda_i-\lambda_j)\frac{\partial^2 S}{\partial\lambda_i\partial\lambda_j},\,\forall i,j=1,\dots,n.
\]
The discussion above should also explain why the coordinates defined in \eqref{MDN-coo}  can be termed  {\em magnetic}. This choice aims at  stressing  the analogy of the present case with the one when to the canonical form of a cotangent bundle is added a {\it magnetic term}, i.e. the pullback of a closed $2$-form defined on the base manifold.}
\end{remark}
Let $(M,N,\omega)$ be a $2n$-dimensional $\omega N$-manifold and let  
\[
\Delta_N(\la)=\la^n-c_1\la^{n-1}-c_2\la^{n-2}-\cdots - c_n
\]
be the minimal polynomial of $N$. 
As it is well known,  the functions $c_k$ and 
$I_k=\displaystyle{\frac{1}{2k}}\operatorname{tr}(N^k)$ are related by the {\em Newton formulas}:
\begin{equation*}\label{eq:3.new}
\begin{split}
&I_1=c_1;\quad I_2=c_2+\frac12 c_1^2;\quad I_3=c_3+c_2c_1+\frac13c_1^3;\\
&I_4=c_4+c_1c_3+c_1^2c_2+\frac12 c_2^2+\frac14 c_1^4;\quad 
I_5=c_5+\ldots\, .\end{split}
\end{equation*}
Using these identities, one can show that the Lenard relations \eqref{eq:lenard} 
imply that the $c_i$'s satisfy the following recursive formulas:
\begin{equation}
  \label{eq:frrecrel}
  N^* dc_i=dc_{i+1}+c_i dc_1,\>\text{ with } c_{n+1}\equiv 0\, ,
\end{equation}
which can be  more compactly written as: 
\begin{equation}
  \label{eq:deltarel}
  N^* d\Delta_N(\la)=\la d\Delta_N(\la)+\Delta_N(\la)d c_1.
\end{equation}
These observations justify the following definition. Let $(M,N,\omega)$ be a $\omega N$-manifold of dimension $2n$ and let $\la_1,\dots,\la_n$ be the eigenvalues of $N$. 
\begin{defi}
A smooth function $\Phi(\la)$ defined on $M$ and depending on a additional parameter $\la$ is called a  {\em \Nij\ function generator}
if there exists a $1$-form $ \alpha_\Phi(\la)$, also depending on $\la$ and regular at $\la=\la_i$ for all $i=1,\dots,n$, such that:
\begin{equation}
  \label{eq:2.eeg}
  N^* d\Phi(\la)=\la d\Phi(\la)+\Delta_N(\la) \alpha_\Phi(\la)\>.
\end{equation}
\end{defi}

One can prove the following

\begin{prop}\label{prop:nfg} 

The $n$-functions $\Phi_i=\Phi(\la_i)$, $i=1,\dots,n$, obtained evaluating the {\it generating}
function $\Phi(\la)$ at $\la=\la_i, i=1,\ldots, n$, are \Nij\ functions,
that is, they satisfy: 
\[
N^* d\Phi_i=\la_i d\Phi_i,\,\forall i=1,\dots,n.
\]
\end{prop}
{\bf Proof.} The proof is simply obtained by the usual chain rule. Let $\Phi\in C^\infty(M)[[\la]]$ be a regular function on $M$, formally depending on the additional parameter $\la$, and let $g$ be another smooth function on $M$ that gives rise to an  {\em evaluation} map
\begin{equation}\label{evmap}
ev_g: C^\infty(M)[[\la]]\to C^\infty(M), \text{  by setting } ev_g(\Phi(x,\la))=\Phi(x,g). 
\end{equation}
If we denote by $d$ the  differential on $M$, one gets
\begin{equation}\label{evdiff}
d(ev_g(\Phi))=d\Phi\big\vert_{\la=g}+\left.\frac{\partial\Phi}{\partial\la}\right\vert_{\la=g} dg.
\end{equation}
Now, let us suppose that $\Phi$ satisfies (\ref{eq:2.eeg})  with a regular $\alpha_\Phi(\la)$, and take $g=\la_j,\, j=1,\dots,n$.
Then, taking (\ref{evdiff}) into account, and using the identities $N^* d\la_j=\la_j d\la_j$ and $\Delta_N(\la_j)=0$ for  $j=1,\dots,n$, one gets:
\begin{eqnarray*}
&&N^*\big(d (ev_{\la_j}(\Phi))\big)
=N^*\Big(d\Phi\big\vert_{\la=\la_j}+\left.\frac{\partial\Phi}{\partial\la}\right\vert_{\la=\la_j} d\la_j\Big)\\
&=&\big[\la d\Phi(\la)+\Delta_N(\la) \alpha_\Phi(\la)\big]\big\vert_{\la=\la_j}+\left.\frac{\partial\Phi}{\partial\la}\right\vert_{\la=\la_j} N^*(d\la_j)\\ 
&=&\la_j\big(d\Phi(\la)\big)\big\vert_{\la=\la_j}+\la_j\left.\frac{\partial\Phi}{\partial\la}\right\vert_{\la=\la_j}  d\la_j\\
&=&\la_j \big(d (ev_{\la_j}(\Phi))\big),
\end{eqnarray*}
\endpf


\begin{remark} 
The minimal polynomial of the \Nij\  tensor of a $\omega N$-manifold is the prototype of a \Nij\ function generator. Its roots, if functionally independent, give half of the \dncoo.
\end{remark}

The relevance of the notion of \Nij\ function generator 
for the search for DN-coordinates stems from the following proposition~(\cite{FP03}). 

\begin{prop}\label{mu-dnf}
 If $\Phi(\la)\in\mathcal N(M)$ is such that 
\begin{equation}
Y\big(\Phi(\la)\big)\vert_{\la=\la_i}=1,\,\forall i=1,\dots,n,\label{eq:fund}
\end{equation}
then $\big(\la_1,\dots,\la_n,\mu_1,\dots,\mu_n\big)$ is a system of (local) DN-coordinates on $M$, where $\mu_i=\Phi(\la_i)$, for all $i=1,\dots,n$.
\end{prop}
{\bf Proof.} Since $\Phi(\lambda)\in\mathcal N(M)$, $N^{\ast}d\mu_i=\la_i d\mu_i,\,\forall i=1,\dots,n$.
 
On the other hand, the condition in \eqref{eq:fund} is equivalent to the fact that $(\la_1,\dots,\la_n,\mu_1,\dots,\mu_n)$ are canonical coordinates, see Remark \ref{rem:imp}.
\endpf

Then we are left to understand when one can find a $\Phi(\la)\in\mathcal N(M)$ fulfilling \eqref{eq:fund}. To this end it is worth to record the following result.

\begin{lem}[\cite{FP03}]
The space $\mathcal N(M)$ of the \Nij\ function generators of $(M,N,\omega)$ is an algebra, which is invariant under the action of the vector field $Y$ defined in \eqref{eq:vech}.
\end{lem}


In this way, starting from a set of  \Nij\ functions generators, one can obtain
further elements of the algebra $\CN(M)$ by repeated applications of the
vector field $Y$.  This observation sometimes can be used to solve the problem of finding a $\Phi(\la)\in\mathcal N(M)$ fulfilling \eqref{eq:fund}, as it is explained in the following proposition.

\begin{prop}[\cite{FP03}]
Let $\Phi\in\mathcal N(M)$ such that
\[
Y^n(\Phi)=\sum_{j=0}^{n-1}a_jY^j(\Phi),
\]
where $Y(a_j)=0$, for all $j=1,\dots,n$. Then Eq. \eqref{eq:fund} can be solved algebraically.
\end{prop}



\section{The CM-system}

In this section we shall describe the bi-Hamiltonian structure of the CM-system,  originally defined in \cite{MM96}, (see also \cite{MCFP03}) 
and, according to \cite{BFMOP10}, we shall show how it can be obtained via a suitable reduction process.
The starting point of the construction of the bi-Hamiltonian structure defined in \cite{MM96} is the observation, due 
to Oshaneltsky and Perelomov, 
that the CM-system admits an extended Lax representation,  i.e. one where the usual Lax representation of the equations of motion
\begin{equation} \label{Laxrep}
 \frac{dL}{dt}=[L,B]
\end{equation}
is supplemented with the ``extension''
\begin{equation}\label{eLaxrep}
 \frac{dX}{dt}=[X,B]+L,
\end{equation}
where:
\[
L_{ij}=p_i\delta_{ij}+\frac{(1-\delta_{ij})}{(x_{i}-x_{j})}, \quad
B_{ij}=\delta_{ij}\sum_{l=1}^n\frac{(1-\delta_{jl})}{(x_j-x_l)^2}-\frac{(1-\delta_{ij})}{(x_i-x_j)^2},\]
$\forall\, i,j=1,\dots, n$ and  $X=\text{diag}\,(x^1,\dots,x^n)$. Before moving on, we record the following definition.
\begin{defi}\label{def:cmat}
A pair of Calogero-Moser $n\times n$-matrices is a pair of matrices $(L,X)$ where 
\[
L_{ij}=p_i\delta_{ij}+\frac{c(1-\delta_{ij})}{(x_{i}-x_{j})},\,\forall i,j=1,\dots, n \quad\text{and}\quad X=\text{diag}\,(x_1,\dots,x_n),
\]
where $c=1$ or $c=\sqrt {-1}$.
\end{defi}
Then on the open subset of $\mathbb R^{2n}$ where the matrix $L$ is semi-simple, with distinct eigenvalues and where the eigenvalues of $L$ and 
the diagonal elements of $X$ are functionally independent, the functions
\[
 I_k=\frac1{k}{\operatorname{tr}L^k}\quad\text{and}\quad J_k=\operatorname{tr}XL^{k-1},\,k=1,\dots,n
 \]
form a system of local coordinates. Furthermore, the brackets $\{\cdot,\cdot\}_0$ and $\{\cdot,\cdot\}_1$, defined by:
\begin{equation}
\begin{array}{l}
\{I_k,I_\ell\}_0=0\ ,\quad \{J_\ell,I_k\}_0=(k+\ell-2)I_{k+\ell-2}\ ,\\
\{J_k,J_\ell\}_0=(\ell-k)J_{k+\ell-2}\ ,\text{but } \{J_1,I_1\}_0=n\, ;\\
\vspace{0.2truecm}
\\
\{I_k,I_\ell\}_1=0\ ,\quad \{J_\ell,I_k\}_1=(k+\ell-1)I_{k+\ell-1}\ ,\\
\{J_k,J_\ell\}_1=(\ell-k)J_{k+\ell-1}\ ,
\end{array}\label{eq:invfunc}
\end{equation}
are Poisson brackets, and they define a bi-Hamiltonian structure for the CM-system, \cite{MM96}.\\
In analogy with the argument used by Kazhdan, Kostant and Sternberg to prove the complete integrability of the CM-system, 
in \cite{BFMOP10} the bi-Hamiltonian structure of the CM-system of \cite{MM96} was recovered via a process of reduction, starting from a 
bi-Hamiltonian structure defined on the cotangent bundle of $\mathfrak{gl}(n)$.
Here, for the reader's convenience, we recall the main points of this construction.

Let $(P_0,P_1)$ be the pair of Poisson tensors on $T^*(\mathfrak{gl}(n))\simeq \mathfrak{gl}(n)\times\mathfrak{gl}(n)$, defined as follows:

\begin{description}
 \item[i)] $P_1$ is the Lie-Poisson
structure associated with the semidirect product $\mathfrak{gl}(n)\rtimes\mathfrak{gl}(n)$ Lie bracket
\begin{equation}\label{semiP1}
 \big[(A_1,B_1),(A_2,B_2)\big]=\big(B_1A_2-B_2A_1,[B_1,B_2]\big).
\end{equation}
\item[ii)] $P_0$ is the Lie derivative of $P_1$ w.r.t. the vector field $(\dot{A},\dot{B})=(\mb{Id}, \mb{0})$, also known as the  
{\em freezing} of $P_1$ at the point $(A,B)=(\mb{Id},\mb{0})$. 
\end{description}
We notice that $P_0$ is nothing but the (inverse of the) 
canonical symplectic structure on $T^*(\mathfrak{gl}(n))$, used in the KKS-theory 
of the CM-system. 
By standard result of bi-Hamiltonian geometry, $P_1$ and $P_0$ are compatible Poisson tensors, and so $( T^*(\mathfrak{gl}(n)), P_0^{-1}, P_1\cdot P_0^{-1})$ 
is a $\omega N$-manifold.

The reduction of the pair $(P_0, P_1)$ to the phase space of the CM-system can be performed in two steps, as follows.

\emph{The first projection.}
Let $G=\operatorname{GL}(n)$ and consider the $G$-action on $M$ defined by the simultaneous conjugation, i.e.
\[
\left(g,(A,B)\right)\mapsto \left(gAg^{-1},gBg^{-1}\right)\ .
\]
A nice quotient is obtained by restricting  the $G$-action to a suitable (open and invariant) subset $\mathscr M\subset M$, see Section $4$ of \cite{BFMOP10}.
Then $\mathscr M/G$ is an open set in $\mathbb{R}^{n^2+1}$.

Since the Hamiltonians $H_k=\frac{1}{k}\tr A^k$ are  invariant with respect to the $G$-action, the vector fields $X_k=-P_0dH_k$, of the bi-Hamiltonian hierarchy, given by $\left(X_k\right)_{(A,B)}=\left(0,A^{k-1}\right)$,
can be projected on $\mathscr M/G$. Their projections are the vector fields associated with the 
Hamiltonians $H_k=\frac{1}{k}\operatorname{tr} A^k$, (seen as functions on the quotient manifold) by the reduced bi-Hamiltonian structure. As shown in ~\cite{BFMOP10}, these projected vector fields acquire the Lax-type form
\[
\partial_k A=[\xi_k, A],\quad \partial_k B=[\xi_k, B]+A^{k-1}.
\]
Also, writing explicitly the \Nij\ tensor $N$ associated with the Poisson pair given by~(\ref{semiP1}) and its freezing, it is immediate to check that the eigenvalues of  $N$ are indeed the eigenvalues of the matrix $A$.

\begin{remark}
The functions $H_k=\frac1{k}\tr A^k$, for $k\ge 1$,
form a {\it Lenard-Magri bi-Hamiltonian hierarchy} on $M$, i.e  
\[
N^* dH_k=dH_{k+1},\,(\Leftrightarrow P_1 dH_k=P_0 dH_{k+1})\, .\]
\end{remark}

\emph{The second projection.} To pass from the $(n^2+1)$-dimensional quotient 
$\mathscr M/G$ to the phase space of the CM-system one can observe that the invariant functions
\[
I_k(A,B)=\frac1{k}\tr A^k=H_k(A,B)\ ,\quad J_k(A,B)=\tr(A^{k-1}B)\ ,\,\,\mbox{for $k=1,\dots,n$,}
\]
form a Poisson sub-algebra with respect to both the Poisson brackets $\{\cdot,\cdot\}_0$ and $\{\cdot,\cdot\}_1$, since, on $\mathscr M/G$, they satisfy the relations \eqref{eq:invfunc}.
Thus both Poisson brackets can be further projected on the quotient space defined by the map 
\[
p:\mathscr M/G\twoheadrightarrow U\subset\mathbb{R}^{2n}
\] 
whose components are, by definition, the functions $(I_1,\dots,I_n,J_1,\dots,J_n)$. 


One can show that, locally and up to coverings, the sub-manifold $\mathcal Q\subset M$ whose elements are the pairs entering the {\it extended} Lax representation of the CM-system (\ref{Laxrep}, \ref{eLaxrep}),  is in 1-to-1 correspondence with an open subset of the second quotient space $U$. Furthermore, the projected (bi-Hamiltonian) flows are those of the Calogero-Moser system.
The bi-Hamiltonian structure so obtained is, by construction, expressed in the coordinates $(I_1,\dots,I_n,J_1,\dots,J_n)$, and it is defined by the brackets in \eqref{eq:invfunc}. 
\begin{rem}
 Our definition of the Poisson pair $(P_1,P_0)$ and namely the choice of $P_1$ (see Eq. (\ref{semiP1})), does not come out of the blue. As shown in  \cite{BFMOP10}, it can be framed within the theory of {\em complete lift} of (torsionless) $(1,1)$ tensors from a manifold to its tangent bundle (see, e.g., \cite{IMM00} and \cite{YI73}).    
\end{rem}
\begin{rem}
It worth to recall, for the reader's convenience,  how the Calogero-Moser matrices $(L,X)$ (\ref{def:cmat}) were obtained in \cite{KKS78}, where
the complete integrability of the CM-system was proved via the Hamiltonian reduction of a linear flow defined on the cotangent bundle of $\mathfrak h(n)$, 
the vector space of the $n\times n$ Hermitian matrices. 
On this symplectic vector space, the action $g.(A,B)\rightsquigarrow (gAg^{-1},gBg^{-1})$ of $U(n)$, 
the group of the $n\times n$ unitary matrices, is Hamiltonian with moment map given by:
\begin{equation*}\label{momentmap}
 \mu(A,B)=[A,B],\qquad \forall A,B\in\mathfrak h(n).
\end{equation*}
In \cite{KKS78}, both the Hamiltonian structure of  the phase-space of the CM-system, 
and the corresponding flows, were recovered 
performing the Hamiltonian reduction  from the level set $\mu^{-1}(\alpha)$, where  
\begin{equation*}\label{mominage}
\alpha=\sqrt{-1}\>\mb{e}^T\otimes\mb{e}\quad \text{with}\quad \mb{e}^T=(1,1,1,\dots, 1). 
\end{equation*}

More precisely, denoting with $G_\alpha$ the stabilizer of $\alpha$ in $U(n)$, in \cite{KKS78} it was shown that on the $G_\alpha$-orbit of any $(A,B)\in\mu^{-1}(\alpha)$, there is a unique pair of Calogero-Moser matrices $(X,L)$, with $c=\sqrt{-1}$. 
This observation allows to identify the set of all pairs of Calogero-Moser matrices of this kind with the reduced phase-space $\mu^{-1}(\alpha)/G_\alpha$, which turns out to be symplectomorphic to (an open subset of) $\mathbb R^{2n}$ with its standard symplectic structure. 
\end{rem}

\section{DN-coordinates for the CM-system}\label{sectSKf}


We will now discuss the problem of how to define a full set of DN-coordinates on the phase-space of CM-system, that is, we will discuss the problem of providing a set of canonical momenta $\mu_k$, conjugated to the eigenvalues $\la_k$ of the Lax matrix $L$ of the CM-system. Hereafter $(L,X)$ will denote a pair of Calogero-Moser matrices.

Let $\Delta(\la)=\det\,(\la-L)$ and let us consider the generating function:
\begin{equation}\label{Ggendef}
\Gr:=\operatorname{tr}(X\cdot (\la-L)^\vee),
\end{equation}
where $A^\vee$ denotes the ``classical" adjoint of the matrix $A$, i.e. the transpose of the cofactor matrix of $A$. 
Then:
\begin{prop}
$\Gr$ is a Nijenhuis function generator.
\end{prop}
{\bf Proof.}
Thanks to Proposition \ref{prop:nfg} it suffices to prove that:
\[
N^*d\Gr\equiv \la d\Gr\quad\text{mod }\Delta(\la).
\]
Since both $\Gr$ and $N^*d\Gr$ are polynomials in $\la$, their poles are located at $\la=\infty$. For this reason it is possible to freely use the  asymptotic formal expansion
\[
\Gr=\Delta(\la){\operatorname{tr}}\big(X\cdot (\la-L)^{-1}\big)=\Delta(\la)\sum_{a=0}^\infty \dsl{\frac{J_{a}}{\la^{a+1}}}\ .
\]
 Equation (\ref{eigenNstar}) implies that $N^* dJ_a=d J_{a+1}+ R_a$,
where, written in terms of the, still unknown, DN-coordinates, $R_a=\sum_{j=1}^n\la_j^a d\mu_j$.
Then, applying $N^*$ to $d\Gr=d\Delta(\la)\sum_{a=0}^\infty \dsl{\frac{J_{a}}{\la^{a+1}}}+
\Delta(\la) \sum_{a=0}^\infty \dsl{\frac{d J_{a}}{\la^{a+1}}}$, one has
\begin{equation}
 N^*d\Gr=\la d\Gr+\Delta(\la)\sum_{a=0}^\infty \frac{R_{a}}{\la^{a+1}},
\end{equation} 
whence the assertion, since $\dsl{\sum_{a=0}^\infty \frac{R_{a}}{\la^{a+1}}}$ is regular for $\la=\la_j$.
\endpf

Let us now identify the conjugate momenta, following the procedure described in Proposition \ref{mu-dnf}. Here and below, we shall need the following simple result.

\begin{lem}\label{lemmino}
 Let us consider $F(L)$ and $G(L,X)$ (that is, $F$ depends only on the first element of the pair $(L,X)$). Then:
 \begin{equation}\label{eqbr0}
 \{F(L), G(X,L)\}_0=\operatorname{tr}\Big(\dpt{F}{L}\cdot\dpt{G}{X}\Big).
   \end{equation} 
\end{lem}
{\bf Proof.} It follows by a direct computation, making use of the definition of the reduced Poisson bracket $\{\cdot,\cdot\}_0$, of the definition of Hamiltonian vector field and, in particular,  of the explicit expression of the symplectic form on $T^*(\mathfrak{gl}(n)$
\[
 \omega_0=\sum_{i,j=1}^ndB_{ij}\wedge dA_{ji}.
\]
\endpf
Let $Y=-P_0(d\sum_i \la_i)=-P_0 d\operatorname{tr}(L)$. Then: 
\begin{equation}
 Y(\Gr)=\{ -\operatorname{tr}(L)\, , \operatorname{tr}X\cdot (\la-L)^\vee \}_0=\operatorname{tr}(\la-L)^\vee
=\frac{d}{d \la}\Delta(\la)\label{eq:y}
 \end{equation}
and, trivially,  
\[Y(Y(\Gr))=0.
\]
This yields $ \dsl{Y\left(\frac{\Gr}{Y(\Gr)}\right)=1}$ and so, using Proposition \ref{mu-dnf}, one deduces that 
\[
\left.\frac{\Gr}{Y(\Gr)}\right\vert_{\la=\la_i}=\mu_i,\,\,\forall i=1,\dots,n
\]
are the momenta canonically conjugated to the eigenvalues of the Lax matrix.

\begin{remark}\label{mark}
\begin{enumerate}
\item Note that some of the computations above were performed on the manifold $\mathscr M/G$ obtained after the first projection. 
This is natural in view of the fact that the Poisson brackets on the Calogero-Moser manifold are obtained from those on $\mathscr M/G$ by projection.
\item A simple way to ascertain that $Y(Y\Gr))=Y(\operatorname{tr}(\la-L)^\vee)=0$ working directly on the fully reduced Calogero-Moser manifold $\mathcal{Q}$ 
is to notice that, since $\operatorname{tr}(L)=\sum_i y_i$, one has $Y=-\sum_i \ddd{}{x^i}$ and $L$ (and thus $(\la-L)^\vee)$ 
depends only on the differences $x_i-x_j$ of the physical locations  of the particles.
\end{enumerate}
\end{remark}

\subsection{Spectral Coordinates and  Sklyanin's formula }
The notion of spectral canonical coordinates lies at the very heart of the algebro-geometrical approach to integrable systems (see, e.g., \cite{VN85, DKN90, AHH93, DD94, Sk95,KP99}). 
When the Lax matrix contains a spectral parameter $z$, Darboux coordinates are given by 
the location of the poles of a suitably normalized eigenvector of the Lax matrix. Equivalently, the coordinates are given by the locations on the spectral curve  $\operatorname{det}(\la-L(z))$ of the points corresponding to the zeros of a specific polynomial $\mathcal{P}(z)$ (e.g., for the stationary flows of the KdV hierarchy, the polynomial is the $[1,2]$ element of the Lax matrix).
This method cannot, however, directly be applied to the rational CM-system, since there is no spectral parameter in its {\it standard} Lax representation. 

A specific formula for canonical spectral coordinates for the 
CM-system was proposed by E.K Sklyanin in a talk in Cambridge in 2009. 
It reads:
\begin{equation}\label{skeq}\begin{split}
&\lambda_i:= \text{roots of } {\det}(\la-L);\\
&\left.\widetilde{\mu}_i={\mathcal{E}(\la)}/\dsl{\frac{d}{ d\la} {\det}(\la-L)}\right\vert_{\la=\la_i}
\end{split}
\end{equation} 
with $\Ek= {\mathbf x}^T \cdot (\la-L)^\vee\cdot {\mathbf e}$, $ {\mathbf x}^T=( x_1,\dots,x_n)$ and ${\mathbf e} =( 1,\dots,1)^T$. 


We shall  discuss and prove this formula in the framework exposed so far, following the setting discussed in, e.g.,~\cite{FP03, FP11}. In such an approach, the bi-Hamiltonian structure plays the a  basic role and induces  
the algebro-geometric structure as an output, while,  in the approach of the Moscow
and Leningrad school, the building block is the algebraic geometry of Riemann surfaces and moduli thereof, and the Poisson structure(s) are an output. The main result of this section is contained in the following proposition.

\begin{prop}\label{propekgf}
Formula (\ref{skeq}) defines a complete set of MDN-coordinates for the pair $\{\cdot,\cdot\}_0$, $\{\cdot,\cdot\}_1$, according to the definitions of Section \ref{mdn}.
\end{prop}
We shall actually prove that a set of canonically conjugated variables to the eigenvalues of the Lax matrix is given by   
\begin{equation}\label{skeqbis}
\widetilde{\mu}_i=\left.\frac{{\mathbf x}^T \cdot (\la-L)^\vee\cdot {\mathbf e}}{{\mathbf e}^T \cdot (\la-L)^\vee\cdot {\mathbf e}}\right\vert_{\la=\la_i}\,i=1,\dots,n\, ,
\end{equation} 
a formula which is equivalent to that is (\ref{skeq}) in view of the equality (which shall be proven as well)
\begin{equation}
{\mathbf e}^T \cdot (\la-L)^\vee\cdot {\mathbf e}=\frac{d}{d \lambda} {\det}(\la-L).
\end{equation}

The proof of Proposition \ref{propekgf} can be split in a few steps. We start with two Lemmas, the first of which parallels the content of Proposition \ref{prop:nfg}. 
\begin{lem}\label{A1}
Let $F(x,\la)\in C^\infty(M)[[\la]]$ a regular function on $M$,  depending on the additional parameter $\la$, and suppose that:
\begin{equation}
\label{aug1}
\{\la_k, F(x,\la)\}_0=0,\> k=1,\ldots, n\quad \left( \Leftrightarrow  \ddd{}{\mu_k} F(x,\la)=0,\>k=1,\ldots, n.\right)
\end{equation}
Then:
\begin{equation}
\label{aug2}
\{\la_k, F(x,\la_j)\}_0=0,\> j,k=1,\ldots, n\quad \left(\Leftrightarrow  \ddd{}{\mu_k} F(x,\la_j)=0,\>j,k=1,\ldots, n\right),
\end{equation}
where $F(x,\la_j)$ is, using the notation of Proposition \ref{prop:nfg}, $ev_{\la_j}(F(x,\la))$.
\end{lem}
{\bf Proof.} It follows from the relation
\begin{equation*}
\label{aug3}
d \left(F(\la_j)\right)=dF(\la)\big\vert_{\la=\la_j}+\left.\frac{\partial F}{\partial\la}\right\vert_{\la=\la_j} d\la_j\, .
\end{equation*}
\endpf

\begin{lem}\label{A2}
Let  $I_k=\dsl{\frac1{k}} \operatorname{tr}(L^k)$. Then, for any $G\in C^\infty(M)$,  
\begin{equation*}
\{\la_k, G\}_0=0,\> \forall k=1,\ldots, n\quad\Leftrightarrow \quad \{I_k, G\}_0=0,\> \forall k=1,\ldots, n.
\end{equation*}
\end{lem}
{\bf Proof.} It follows from the fact that, since $I_k=\frac{1}{k}\sum_{i=1}^n \la_i^k$,
the $d\la_k$'s are related to the $dI_\ell$'s by the invertible Vandermonde matrix $VdM(\la_1,\ldots, \la_n)$. 
\\
\endpf
{\bf Proof of Proposition  \ref{propekgf}.} Consider  $\Ek={\mathbf x}^T \cdot (\la-L)^\vee\cdot {\mathbf e}$ and recall 
the definition   $\Gr=\operatorname{tr}(X\cdot (\la-L)^\vee)$. To prove the proposition one must  show that 
\[
\widetilde{\mu}_i-\mu_i=f_i(\la_1,\dots,\la_n),\,\forall i=1,\dots,n,
\]
where the $f_i$'s are suitable smooth functions.  One first shows that 
\begin{equation}\label{aug4}
\ddd{}{\mu_k}\left({\mathcal{E}}(\la_i)-{\mathcal{G}}(\la_i)\right)=0,\,\forall i,k=1,\dots,n.
\end{equation}
In view of the two lemmas above, the proof of the previous formula boils down to check that, for all $k=1,\ldots, n$, 
\begin{equation}\label{1-7.1}
 \{{\mathcal{I}(\la)}_k, \Ek-\Gr\}_0=0,
\end{equation} 
where ${\mathcal{I}(\la)}_k=\dsl{\frac{1}{k}\operatorname{tr}(\la-L)^k}$.
To ascertain the validity of such a relation one simply rewrites the Sklyanin generator $\Ek$ as:
\begin{equation*}
 \Ek={\mathbf x}^T \cdot (\la-L)^\vee\cdot {\mathbf e}=
 \mb{e}^T\cdot  X \cdot (\la-L)^\vee\cdot \mb{e}=
 \operatorname{tr}\big( X\cdot (\la-L)^\vee \cdot({\mathbf e}\otimes{\mathbf e}^T)\big),
\end{equation*} 
so that one can compactly write
\begin{equation*}
 \Ek-\Gr=\operatorname{tr}\Big(X\cdot(\la-L)^\vee \cdot \big(\mathbf{e} \otimes{\mathbf e}^T-\mathbf{Id}_n\big)\Big).
\end{equation*} 
Using Lemma \ref{lemmino}, one gets:
\begin{eqnarray*}
 &&\{{\mathcal{I}(\la)}_k, \Ek-\Gr\}_0=
  \operatorname{tr}\Big(\dpt{{\mathcal{I}(\la)}_k}{L}\cdot\dpt{\big(\Ek-\Gr\big)}{X}\Big)\\
  &=&\operatorname{tr} \big((L-\la)^{k-1}\cdot (\la-L)^\vee \cdot (\mathbf{e} \otimes{\mathbf e}^T-\mathbf{Id}_n)\big).
\end{eqnarray*} 
The proof of \eqref{aug4} follows at once using the characteristic Calogero formula, see Equation (\ref{mominage}),
\begin{equation*}
 \big(\mathbf{e} \otimes{\mathbf e}^T-\mathbf{Id}_n\big)
 =[\la-L,X],
\end{equation*} 
and recalling that any matrix commutes with its classical adjoint.

To conclude the proof of the proposition, now it suffices to show that 
\[
{\mathbf e}^T \cdot (\la-L)^\vee\cdot {\mathbf e}= Y(\Gr).
\]
To this end, first observe that, since $Y$ is the Hamiltonian vector field associated with $-\operatorname{tr}(L)$, the specialization of Equation (\ref{1-7.1})  to  $k=1$ yields
\[
 Y(\Ek)=Y(\Gr).
\]
On the other hand (see Remark \ref{mark}) by an explicit computation one shows that
\[
 Y(\Ek)={\mathbf e}^T\cdot (\la-L)^\vee\cdot {\mathbf e}\>,
\]
and, respectively, that
\begin{equation*}\label{YG}
Y(\Gr)=Y\, (\operatorname{tr}(X\cdot (\la-L)^\vee))= \operatorname{tr}((\la-L)^\vee)=\frac{d}{d \la}\operatorname{det}(\la-L).
\end{equation*}
\endpf

\begin{remark}
We have checked up to $n=5$ that the relation between the generators $\Ek$ and $\Gr$ is the following simple one:
\begin{equation*}
\label{aug5}
\Ek=\Gr+\frac12 \frac{d^2}{d\, \la^2} {\det}(\la-L)\, .
\end{equation*}
We conjecture this to hold for all $n$'s.
\end{remark}

\newpage
\subsection{A few applications}

\subsubsection{The additional integrals.}
As it is well-known, the CM Hamiltonian $H_{CM}=I_2$ 
is {\em super--integrable}~\cite{Wo83}, that is, it admits, beyond the mutually commuting
integral $I_j$, further $n-1$ integrals $\Gamma_\ell, \>
\ell=1,3,4,\cdots,n$. 
Let us  recover them in this formalism.

From the commutation relations (\ref{eq:invfunc}) we get, setting $\ell=2$,
\begin{equation*}
\label{H2pb}
\{I_k,I_2\}_0=0\ ,\quad \{J_\ell,I_2\}_0=\ell I_{\ell}.
\end{equation*}
Consider
\[
  \Gamma_{\ell}=\ell J_2I_\ell-2J_\ell I_2.
\]
It is easy to show that $\langle I_i, \Gamma_k\rangle_{i,k=1,\ldots n,\> k\neq 2}$
are well defined and functionally independent, and they satisfy
\[
\{\Gamma_{\ell},I_2\}=0.
\]
\subsubsection{The asymptotic positions and momenta.}
In the repulsive case, as discussed first in \cite{Mo75}, it was remarked that the time-asymptotic formula for the solutions of the CM-system is
\begin{equation*}
\label{aug7}
\left\{\begin{array}{lcc}
x_k(t)=p_k^\pm t+x_{k}^\pm+o(t^{-1}) && t\to\pm\infty\\
p_k(t)=p_k^\pm\, .&&
\end{array}
\right.
\end{equation*}
It is interesting to note that in this asymptotic regime, that is, when $|x_k-x_\ell|\to \infty$, since the Lax matrix tends to the diagonal matrix
\[
L^\pm=\left(
\begin{array}{cccc}
p_1^\pm&&&\\
&p_2^\pm&&\\
&&\ddots&\\
&&&p_N^\pm
\end{array}
\right)
\]
one gets that $\Ek\to\Gr$, and, furthermore, in the asymptotic limit $|x_k-x_\ell|\to \infty$:
\begin{equation}
\label{aug6}
\mu_k\simeq\widetilde\mu_k\to - x_k^\pm,\,\,\forall k=1,\dots,n,
\end{equation}
see \cite{Pebook} p.172.\\

\subsubsection{The ``eigenvector'' formula.}
As a final
simple application of  Sklyanin's formula, let us show how we can recover the so-called ``eigenvector" formula, 
that is   the compact formula giving the $\mu_i$'s as 
\[
\widetilde{\mu}_i=\frac{\mathbf{x}^T\cdot \bs{\Psi}_i}{ {\mathbf e}^T\cdot \bs{\Psi}_i} \quad \text{
where }\> L \cdot \bs{\Psi}_i=\la_i \bs{\Psi}_i\>.
\]
In our formalism we re-write
\[
\widetilde{\mu}_i=\left.
\frac{{\mathbf x}^T\cdot  (\la-L)^\vee \cdot {\mathbf e}}
{ {\mathbf e}^T (\la-L)^\vee \cdot {\mathbf e}}\right\vert_{\la=\la_i},
\]
and simply  notice that if $A$ is a matrix with one-dimensional 
kernel spanned by $\bs{\Psi}$, then its adjoint is the rank one matrix given by
\[
A^\vee =\bs{\Psi}\otimes \bs{ \Psi}^{*,T}\,,
\]
where $\bs{\Psi}^{*,T}$ is 
a suitable generator of the left kernel of $A$. 

This shows how the MDN-coordinates defined in this paper fit  the standard theory of the canonical  coordinates associated  with the Lax spectral problem (see, e.g., \cite{DKN90}).

\subsection*{Acknowledgments}  GF wishes to thank the ICMC-USP at S\~{a}o Carlos, where part of the work was carried out, as well as the financial support of FAPESP, process number 2014/03665-6. Also, GF thanks B. Dubrovin and F. Magri for discussion about this subject. This work was carried out under the auspices of the GNFM Section
of INdAM. Partial support  the MIUR PRIN project 2010JJ4KPA {\em
Geometric and analytic theory of Hamiltonian systems in finite and infinite dimensions}
is acknowledged.

\end{document}